\documentclass[preprint,12pt]{aastex}
\def\ltsima{$\; \buildrel < \over \sim \;$}
\def\simlt{\lower.5ex\hbox{\ltsima}}   
\def\gtsima{$\; \buildrel > \over \sim \;$}
\def\simgt{\lower.5ex\hbox{\gtsima}}
\newcommand\bcite[1]{\citeauthor{#1} \citeyear{#1}}

\def\gammin{\gamma_{\rm min}}
\def\btheta{\hbox{\boldmath{$\theta$}}_{\hbox{\tiny{I}}}}
\def\thp{\hbox{\boldmath{$\theta$}}'}
\def\bbeta{\hbox{\boldmath{$\theta$}}_{\hbox{\tiny{S}}}}
\def\zl{z_{\hbox{\tiny{L}}}} \def\zs{z_{\hbox{\tiny{S}}}}
\def\fls{c_{\hbox{\tiny{LS}}}}
\def\Dl{d_{\hbox{\tiny{OL}}}} \def\Ds{d_{\hbox{\tiny{OS}}}}
\def\Dls{d_{\hbox{\tiny{LS}}}}
\def\Gyr{\,{\rm Gyr}}
\def\legacy{\,{\rm km\,s^{-1}\,Mpc^{-1}}}

\begin{document}

\title{Radial density profiles of time-delay lensing galaxies}

\author{J. I. Read$^1$, P. Saha$^1$, and A. V. Macci\`o$^{1,2}$}

\affil{$^1$Institute of Theoretical Physics, University of Z\"urich, \\
Winterthurerstrasse 190, 8057 Z\"urich, Switzerland \\
$^2$Max-Planck-Institut f\"ur Astronomie,
K\"onigstuhl 17, 69117 Heidelberg, Germany}

\date{}

\begin{abstract}

We present non-parametric radial mass profiles for ten QSO strong
lensing galaxies.  Five of the galaxies have profiles close to
$\rho(r)\propto r^{-2}$, while the rest are closer to $r^{-1}$, consistent with an NFW profile.  The former are all relatively isolated early-types and dominated by their stellar light. The latter ---though the modeling code did not know this--- are either in clusters, or have very high mass-to-light, suggesting
dark-matter dominant lenses (one is a actually pair of merging galaxies).  The same models
 give $H_0^{-1} = 15.2_{-1.7}^{+2.5}\Gyr$ ($H_0 = 64_{-9}^{+8} \legacy$), consistent with 
a previous determination.  When tested on simulated lenses taken
from a cosmological hydrodynamical simulation, our modeling pipeline
recovers both $H_0$ and $\rho(r)$ within estimated uncertainties.

Our result is contrary to some recent claims that lensing time delays
imply either a low $H_0$ or galaxy profiles much steeper than
$r^{-2}$.  We diagnose these claims as resulting from an invalid
modeling approximation: that small deviations from a power-law profile
have a small effect on lensing time-delays.  In fact, as we show using
using both perturbation theory and numerical computation from a
galaxy-formation simulation, a first-order perturbation of an
isothermal lens can produce a zeroth-order change in the time delays.

\end{abstract}


\maketitle

\section{Introduction}\label{sec:introduction}

In our current understanding of structure formation, galaxies form
from the dissipation of gas within dark matter halos \citep[see 
e.g.,][]{1978MNRAS.183..341W}. Cosmological simulations suggest that these halos are
self-similar, with universal density profiles
\citep[see
e.g.,][]{1996ApJ...462..563N,2004MNRAS.353..624D,2004MNRAS.355..794H,
2006AJ....132.2685M,2006AJ....132.2701G,2006AJ....132.2711G, 1998ApJ...499L...5M}.
The dark-matter halos are commonly fit to the well-known NFW profile
\begin{equation}
\rho \propto \frac{1}{(r/a)^\alpha(1+r/a)^{3-\alpha}}
\label{eqn:rhosp}
\end{equation}
with $\alpha$ is numerically found to be in the range $\alpha \sim [1,1.5]$, but other
parameterizations are also possible
\citep[e.g.,][]{2006AJ....132.2685M,2006AJ....132.2701G,2006AJ....132.2711G}.
Over the visible region of galaxies ($r \ll a$), the dark matter distribution tends towards a single power law with: $\rho \sim r^{-\alpha}$. At much larger radii ($r \gg a$), the distribution tends towards $\rho \sim r^{-3}$.

While the distribution of dark matter is now relatively well understood, less clear is the effect of the gaseous component on the final {\it
total} mass distribution of galaxies, since simulations involving gas
remain a significant technical challenge \citep[see
e.g.,][]{2004bdmh.conf...37M}. However, it is likely to be more
complex than a simple sum of the gaseous, stellar and dark matter
components, since the gas collapse will cause a contraction of the
underlying dark matter distribution, increasing the central
concentration of dark matter, and possibly even making the halo more
spherical \citep[e.g.,][]{1980ApJ...242.1232Y,2004ApJ...611L..73K}.
Despite these worries, the final resulting mass distribution is
simpler to predict if the stars and gas form the dominant dynamical
component of the galaxy -- as is the case in the centre of massive
spiral galaxies and ellipticals. There we may expect an isothermal
distribution ($\rho \propto r^{-2}$), either as a result of
equilibrium gas physics \citep{1991pagd.book.....S}, or relaxation 
\citep{1967MNRAS.136..101L}. Indeed, dynamical measurements of massive
galaxies suggest that an isothermal distribution provides an excellent
fit over a wide range of radii \citep[see e.g.,][]
{1991MNRAS.253..710V,2000AAS..144...53K,2006ApJ...646..899H}.  At
larger radii, however, we expect a transition from $\rho \sim r^{-2}$ to $\rho \sim r^{-1}$ as the dark matter halo becomes more and more dynamically dominant. Eventually, at very large radii ($r \gg a$) we expect a second transition to $\rho \sim r^{-3}$.

In this paper, we test the above theoretical predictions by using
strong lensing to determine the non-parametric density profile of ten
QSO lensing galaxies, for the first time. Strong lensing measurements, which are limited in radius to the outermost observed image, probe radii $r < a$. Thus, we hope to probe the transition from $\rho \sim r^{-2}$ to $\rho \sim r^{-1}$ as we move from galaxies dynamically dominated by their stars, to galaxies dynamically dominated by their dark matter. We do not measure far enough out ($r >a$) to test the prediction of an eventual transition to $\rho \sim r^{-3}$.

\section{The method}\label{sec:method}

\subsection{Searching lens models}\label{sec:searching}

A given set of lensing observables, even with zero noise, is
consistent with a variety of lensing mass distributions.\footnote{See
the Appendix for brief derivation of this and some other relevant
results from lensing theory.}  This is the well-known problem of
lensing degeneracies.  If multiple source redshifts are available, as
is the case in rich lensing clusters, the effect of degeneracies is
minimal, enabling robust mapping of the mass
profiles \citep{2006ApJ...652L...5S}.  For galaxy lenses, however,
multiple source redshifts are very unlikely, and as a result
degeneracies present a serious difficulty.  The most important
degeneracy couples time delays and the steepness of the mass profile:
steeper mass profiles have higher values of $(H_0\,\Delta t)$ for the
same image positions and magnification ratios.

Given the steepness degeneracy, one might hope that if $\Delta t$ and
$H_0$ are both known, the steepness is constrained.  But even that is
not guaranteed, because time delays are also influenced by, for
example, twisting ellipticity
\citep{1999AJ....118...14B,2003ApJ...582....2Z,2006ApJ...653..936S}
implying further degeneracies.  So it would appear that lensing
observables constrain only some dreadful function of $H_0$, steepness,
and shape.

However, the situation is not so hopeless.  The data may indeed allow
a wide range of models, but by searching through the allowed
`model-space' and adding some prior information, one can make
probabilistic inferences.  To do so, one needs (1)~a prior on
model-space, and (2)~an algorithm for sampling allowed models.

The technique of pixelated lens models is one possible strategy for
(1) and (2).  The lens model is constructed as a superposition of mass
pixels.  It turns out that the data can be encoded as linear equations
on the mass pixels --- though there are many more equations than
there are pixels.  Conservative but reasonable priors can be encoded as
linear inequalities on the mass pixels.  Specifically, in this work we
require the projected density to (i)~be non-negative, (ii)~be
inversion symmetric, meaning $\Sigma(-\btheta)=\Sigma(\btheta)$,
unless the galaxy is a known irregular, (iii)~be centrally
concentrated, with the local density gradient pointing at most
$45^\circ$ away from the center, (iv)~have no pixel more than twice
the sum of its neighbors, the central pixel excepted, and (v)~have
circularly averaged profile nowhere shallower than $R^{-\gammin}$ -- we will return to this constraint, and our choice of $\gammin$ in a moment.  The explicit equations and
inequalities are given in early papers \citep{1997MNRAS.292..148S,1998MNRAS.294..734A,1998AJ....116.1541A} for weak as well as strong lensing.  \cite{2000AJ....119..439W} then
introduced the idea of uniformly sampling the model-space that
satisfies all the equations and inequalities, resulting in a
model-ensemble from which estimates and uncertainties can be derived.

The above ideas are implemented in the {\em PixeLens\/} code, which is
described in detail, along with further justification of the prior, in
\cite{2004AJ....127.2604S}. Some later improvements, including
multithreading, are noted in \citeauthor{2006ApJ...650L..17S}
(\citeyear{2006ApJ...650L..17S}); hereafter Paper~I.   Still later
improvements improve the statistical sampling in the ensemble.
Consequently, the ensembles of 100 models which we use in this paper
are actually better sampled that the ensembles of 200 in Paper~I
because they eliminate clusters of strongly correlated
models\footnote{These various developments will be described in
detail in Coles (in preparation).  See {\tt
http://www.qgd.uzh.ch/programs/pixelens/} for the program itself.}.
Several different tests have been done:
\begin{enumerate}
\item recovering $H_0$ from mock data derived from simple parametrized
lenses \citep{2000AJ....119..439W};
\item tests for biases in the model-sampling procedure
\citep{2006A&A...450..461S};
\item comparing the inferred distributions of the dimensionless time delay
$\varphi$ [see Equation~\ref{varphi}] for observed versus simulated
lenses (Paper~I).
\end{enumerate}
The technique does well in all of these tests, notwithstanding the
seeming crudeness of the prior.  It also correctly predicts the
morphology of Einstein rings \citep{2001AJ....122..585S}.  Later in
this Section we show a further test, for simultaneous recovery of both
$H_0$ and the radial mass profile.

Related ideas appear in other work: \cite{2003ApJ...590...39K} used an
ensemble of parametric models to interpret a rare quintuple lens,
whereas \cite{2005A&A...437...39B} and \cite{2005MNRAS.362.1247D} used
free-form lenses but not model-ensembles.

As in Paper I, we do not use flux ratio measurements. Tensor magnifications from fluxes and VLBI constrain time delays between {\it nearby images,} but remarkably, 
they have no discernible effect on longer time delays \citep{2003AJ....126...29R}.
The physical reason is not hard to appreciate: magnification 
is essentially the second derivative of the arrival time surface (see Appendix I). Time delays between widely separated images tend to 
wash out the sort of local variations of density that cause 
differences in magnifications.  The use of tensor magnifications to resolve
substructure with pixelated models is studied in \cite{saha-2007}.

\subsection{The ten-lens model-ensemble}

Paper~I presented $H_0$ derived from ten time-delay lenses: the quads
J0911+055, B1608+656, and B1115+080, and the doubles B0957+561,
B1104--181, B1520+530, B2149--274, B1600+434, J0951+263, and
B0218+357.  The results derived from an ensemble of 200 composite
models.  Each composite model consisted of pixelated mass maps of all ten lenses, with a shared $H_0$.  The value of $H_0$ varied across the ensemble (as shown in the histogram in Figure~1 of Paper~I), leading
to an estimate with uncertainties.  But within each composite model
$H_0$ was the same, thus coupling the data from different lenses.

The coupling of different time-delay lenses through the shared Hubble
parameter is key to constraining both the lenses and $H_0$ in spite of
degeneracies.  Suppose we take from the ensemble a ten-lens model with
a common $H_0$.  Then we make all the lenses steeper, while increasing
$H_0$, as permitted by the steepness degeneracy.  We cannot do so
indefinitely, because the steepness transformation (see
Equation~\ref{rescale}) will eventually create a negative density in
one of the lenses.  Once we reach this point, we cannot increase $H_0$
any more, nor can we make {\em any\/} of the lenses any steeper.  At
the other end, the prior assumes, conservatively, that the lenses are no shallower than
$\Sigma(R)\sim R^{-\gammin}$.  This means that if {\em one\/} lens is at
$R^{-\gammin}$, $H_0$ cannot get any lower and {\em none\/} of the lenses
can get any shallower, even if they are nowhere near $R^{-\gammin}$.  The
full picture is more complicated because of shape degeneracies and the
details of the prior, but the coupling of different lenses through the
shared Hubble parameter does indeed allow both $H_0$ and lens profiles
to be simultaneously constrained.

In previous papers on time delay lensing galaxies (like Paper I), we
used $\gammin = 0.5$. Here, where we have a heterogeneous sample of
galaxies, and where we would like to probe shallower density
distributions, it is not clear that this is the right prior to use.
Therefore we experimented with changing this prior in the range
$\gammin = [0,0.5]$.  We start with the same prior as in Paper I:
$\gammin = 0.5$. Then, we re-run the analysis for the full ensemble,
but for any galaxy which had a density profile at or near $\gammin =
0.5$, we reduce the prior to $\gammin = 0$. This allows shallower
galaxies to be as shallow as they like, without systematically biasing
the sample space available to the steeper ones. Because of the weaker
prior we use in this paper, we include a new determination of $H_0$ in
Figure \ref{fig:hubble}. It is somewhat lower, but consistent within
uncertainties, with the results from Paper I.

\subsection{Deprojection}\label{sec:deprojection}

Starting with the model ensemble, as detailed above, we then circularly average each projected-mass map to obtain a $\Sigma(R)$ and deproject by numerically solving the usual Abel
integral equation:
\begin{equation}
\rho(r) = -\frac{1}{\pi}\int_r^{\infty}
          \frac{d\Sigma(R)}{dR}\frac{dR}{\sqrt{R^2-r^2}} .
\end{equation}
To evaluate the numerical derivative and then the integral, we
linearly interpolate $\Sigma(R)$ up to the $R$ of the outermost image and
assume $\Sigma \propto R^{-2}$ outside.  Since we have an ensemble of
models, we automatically derive uncertainties on $\rho(r)$ as well.
The technique is the same as in \cite{2006ApJ...652L...5S}, and the
result is not sensitive to the assumed outer slope, or the $R$ where
that outer slope begins, provided the latter is not beyond the image
region. 

The above procedure assumes spherical symmetry and a possible concern
is that this may introduce significant systematic error. To test for
this, we projected and then deprojected (using the above method) a
triaxial halo taken from a cosmological $N$-body simulation. We
recovered the {\it spherically averaged\/} density distribution
perfectly, within the errors. This demonstrates that our assumption of
spherical symmetry in the deprojection algorithm is equivalent to
spherically averaging a (mildly) triaxial halo --- a practise which is
already common in the simulation community. It is important to
emphasise that we {\it do not\/} assume spherical symmetry in our mass
model at any stage until this final deprojection.

\subsection{Test of the method}\label{methodtest}

We have tested the simultaneous recovery of $\rho(r)$ and $H_0$ using
lenses derived from a galaxy-formation simulation.  A test against a
range of simulated galaxies, mimicking observed surveys in detail, is
desirable.  Unfortunately, galaxy simulations at the required
resolution and including both dark-matter and gas dynamics are not yet
numerous.  So we simply generate several lenses out of one simulated
galaxy.

The simulated galaxy is taken from a hydrodynamical cosmological
simulation by \cite{2006MNRAS.366.1529M}.  The galaxy is an E1 or E2
triaxial elliptical dominated by stars in the inner region, but with
overall $\sim 80$\% dark matter \cite{2006MNRAS.366.1529M}.  By
ray-tracing through this galaxy, as described in
\citep{2005MNRAS.361.1250M}, we generated five mock lenses: three
doubles and two quads.  The image positions and time delays were fed
into {\em PixeLens,} which then generated a model ensemble consisting
of mass maps and inferred $H_0$-values. We experimented with changing
the minimum steepness constraint: $R^{-\gammin}$ in the prior (see
section \ref{sec:searching}, for a full description of our prior); the
results were insensitive to changes in the range: $\gammin =
[0,0.5]$. In the following tests, we show results for $\gammin = 0$ --
the weaker prior.

\begin{figure}[t]
\begin{center}
\includegraphics[width=0.4\textwidth]{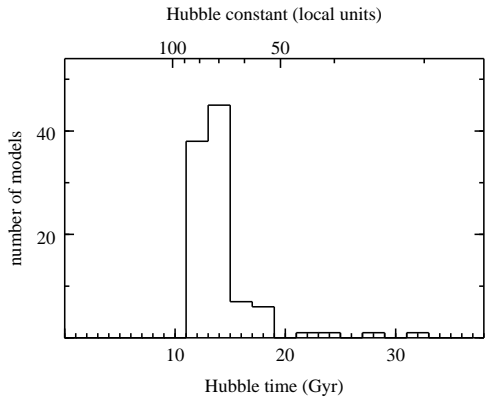}
\caption{The simulated $H_0^{-1}$ as recovered from the model
ensemble.  The unbinned values give $13.6_{-1.4}^{+1.5}\Gyr$. The
correct value was $14\Gyr$.}
\label{fig:nbodyhubble}
\end{center}
\end{figure}

Figure~\ref{fig:nbodyhubble} shows the distribution of $H_0^{-1}$ from
the model ensemble in this test.  We see that the correct value is
well within the 68\%-confidence region.

\begin{figure*}[t]
\begin{center}
\includegraphics[width=\textwidth]{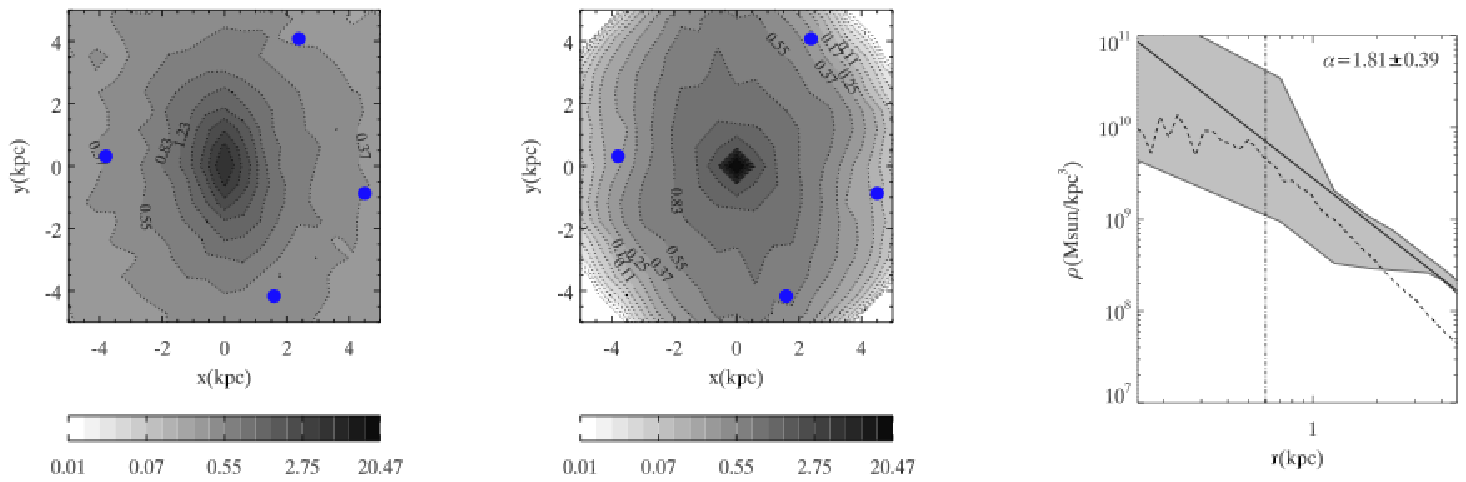}
\caption{Recovered density distribution of the simulated galaxy,
derived from the {\em PixeLens\/} mass map of one of the quads. {\it
Left}: projected mass map for the simulated galaxy; the units for
the contours are in units of the critical density. {\it Middle}:
recovered projected mass map from {\em PixeLens\/} (see text for
details). The solid circles show the distribution of (fake)
images. {\it Right}: deprojected radial density distribution. The
shaded gray region is the 68\% confidence region.  The oblique solid
line is the best-fit power-law $r^{-\alpha}$.  The dashed curve is
the actual spherically-averaged $\rho(r)$.  The plot is truncated at the projected radius of the outermost lens image.  The vertical dot-dashed line to the left marks the
resolution limit of the simulation---the apparent core to the left
of this line is a result of the force softening and is not
physical.}
\label{fig:nbodyprofile}
\end{center}
\end{figure*}

Figure~\ref{fig:nbodyprofile}, left and middle panels, show the 2D projected mass map of the simulated galaxy and the recovered distribution for one of the quads, respectively. Recall that all five mock lenses
had the same mass map, but our modeling codes did not know that; the results from the other four galaxies were nearly identical to those presented here, and we omit these for brevity.

Notice how well the projected mass map is recovered. We obtain the correct elongation along the $y$-direction, while the contours of equal mass density are well-matched within the errors. Beyond the images (where we have no data) the mass distribution drops sharply and is not to be trusted; we do not use any data beyond the outer-most image in our deprojection. 

\begin{figure}[t]
\begin{center}
\includegraphics[width=0.4\textwidth]{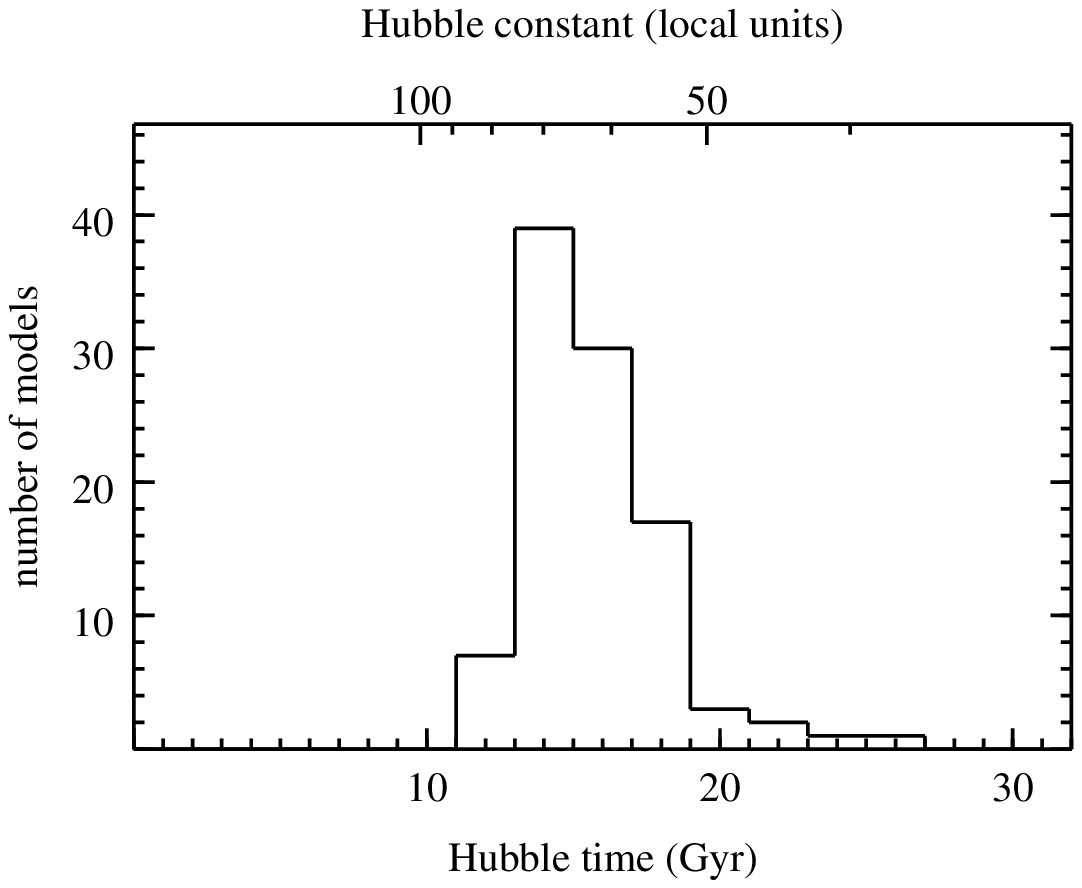}
\caption{$H_0^{-1}$ as recovered from the ten lens
ensemble.  The unbinned values give $15.2_{-1.7}^{+2.5}\Gyr$ ($H_0 =  64.3_{-9}^{+8} \legacy$).}
\label{fig:hubble}
\end{center}
\end{figure}

It is worth commenting a little on this plot. The mass map shown is the average of the full model ensemble; it represents the expectation value of the mass in each pixel. This does not mean, however, that it is not a genuine representation of the projected mass distribution. In this specific test here, {\it PixeLens} accurately recovers the input morphology and mass, within estimated uncertainties. In other similar examples in the literature, {\it PixeLens} has resolved substructure in lensing galaxies and clusters \citep{saha-2007}, and in one example, resolved an interacting pair of galaxies (see Paper I). 

The deprojection of the recovered mass map is shown in the right panel of Figure \ref{fig:nbodyprofile}. The simulated galaxy density profile (dashed line) is well recovered within the errors (grey band). Over-plotted is a power-law fit to the recovered density distribution, obtained using a standard Levenberg-Marquardt least-squares fitting technique (see e.g. \bcite{1992nrca.book.....P}). Note that, although this deprojection assumes spherical symmetry, our mass model up until this point did not (see also section \ref{sec:deprojection}). 

\section{Profiles of observed time-delay lenses}\label{sec:obs}

Having tested the pipeline of recovering both $H_0$ and $\rho(r)$ on
mock lenses generated from a recent galaxy-formation simulation, we
now derive radial profiles for real lenses.

\begin{figure*}
\begin{center}
\includegraphics[width=0.7\textwidth]{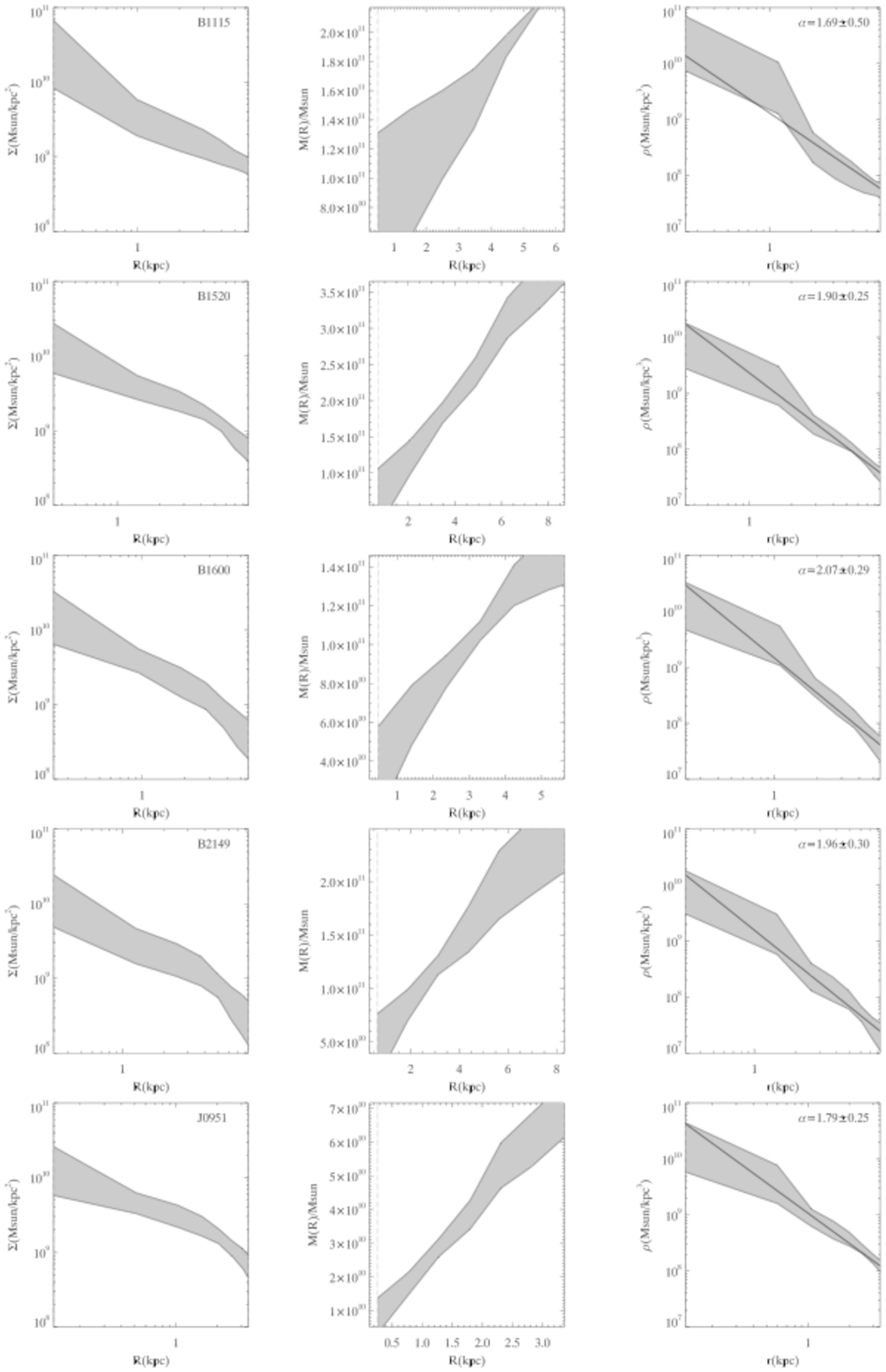} 
\caption{Inferred radial profiles for the lenses. The left column
shows the projected density $\Sigma(R)$. The middle column shows the
enclosed projected mass $M(R)$.  The right column shows the
deprojected density profiles $\rho(r)$ along with power-law fits
$r^{-\alpha}$.  The plots are truncated at the innermost and outermost image
in each system. The gray bands are 68\% uncertainties derived from the
pixelated model ensembles and are probably a good representation of
the uncertainties.  The quoted uncertainties on $\alpha$ are formal
1$\sigma$ errors.  These galaxies are all close to $\rho(r)\propto
r^{-2}$. Continued in Figure~\ref{fig:obs2}.}
\label{fig:obs1}
\end{center}
\end{figure*}

The first part has already been done in \cite{2006ApJ...650L..17S} (Paper I),
where a model-ensemble for ten time-delay lenses was generated, to
estimate $H_0$.  However, in this work we use a weaker prior and so it is worth plotting our determination of $H_0$ again. This is shown in Figure \ref{fig:hubble}; the results are consistent with Paper I, and we recover $H_0^{-1} = 15.2^{+2.5}_{-1.7}\Gyr$ ($H_0 =  64.3_{-9}^{+8} \legacy$). Next, we derive $\rho(r)$, with uncertainties, from the model ensemble, as explained above.  Finally we fit a power law $r^{-\alpha}$ to the non-parametric $\rho(r)$, as in section \ref{methodtest}, above.  Such a simple least-squares power law fit does not do justice to the full ensemble mass distribution which our non-parametric method derives for each galaxy. We use it just as a convenient way of representing our results on a single plot (Figure \ref{fig:galenviron}), and as a simple way to compare our results with theoretical expectations. Our detailed results are shown in Figures \ref{fig:obs1} and \ref{fig:obs2}.

\begin{figure*}
\begin{center}
\includegraphics[width=0.7\textwidth]{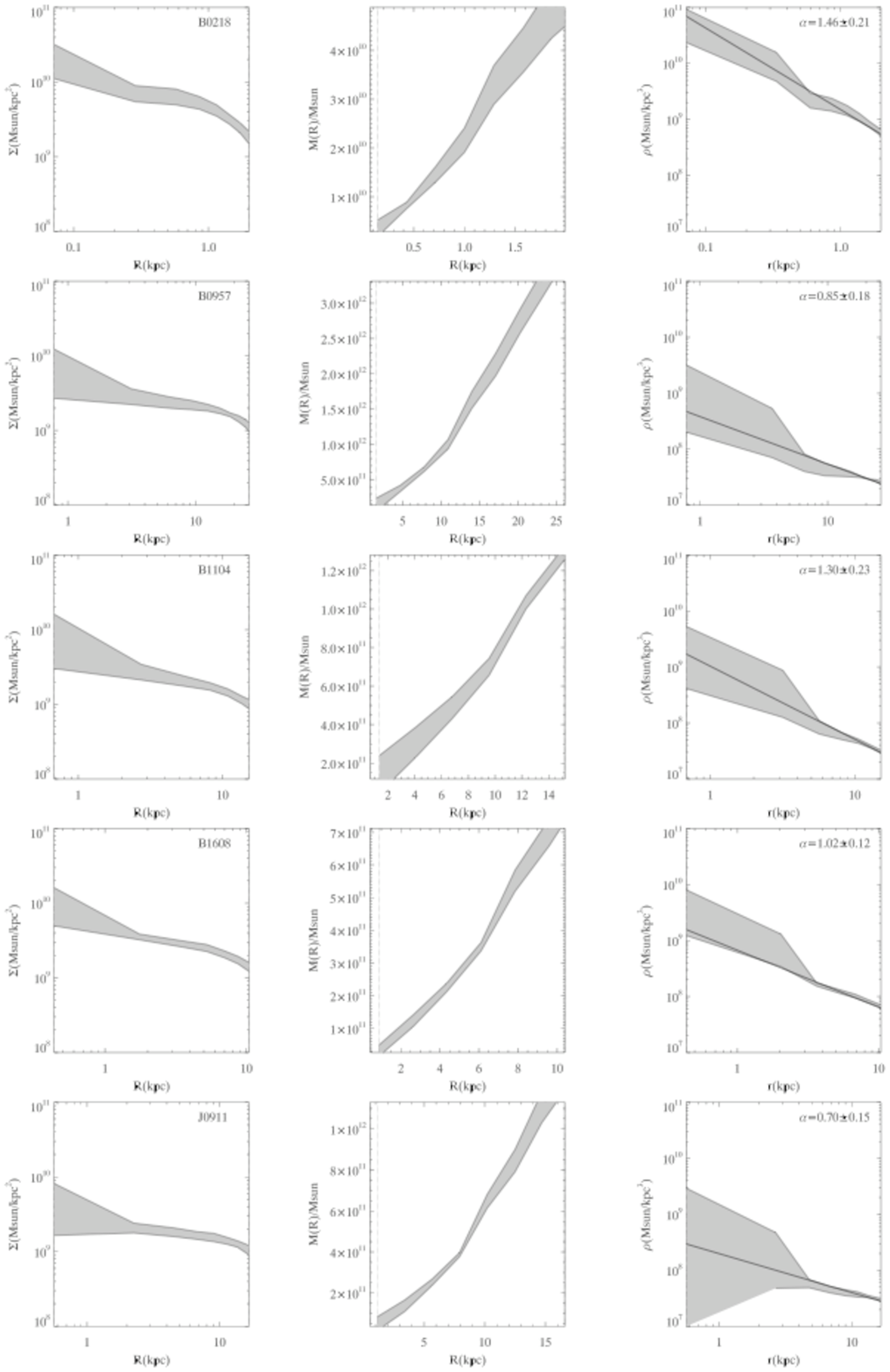} 
\caption{Continuation of Figure \ref{fig:obs1}. The lenses shown here
are all significantly shallower than $r^{-2}$.}
\label{fig:obs2}
\end{center}
\end{figure*}

Figure~\ref{fig:galenviron} then summarizes the log density slopes,
$\alpha$, for all ten lenses.  Five are close to $r^{-2}$ or slightly
steeper, and five are shallower, clustering around $\sim r^{-1}$.
Formal errors on $\alpha$ are shown by the error bars.  These probably
somewhat underestimate the actual uncertainties.

\begin{figure}[t]
\begin{center}
\includegraphics[width=0.49\textwidth]{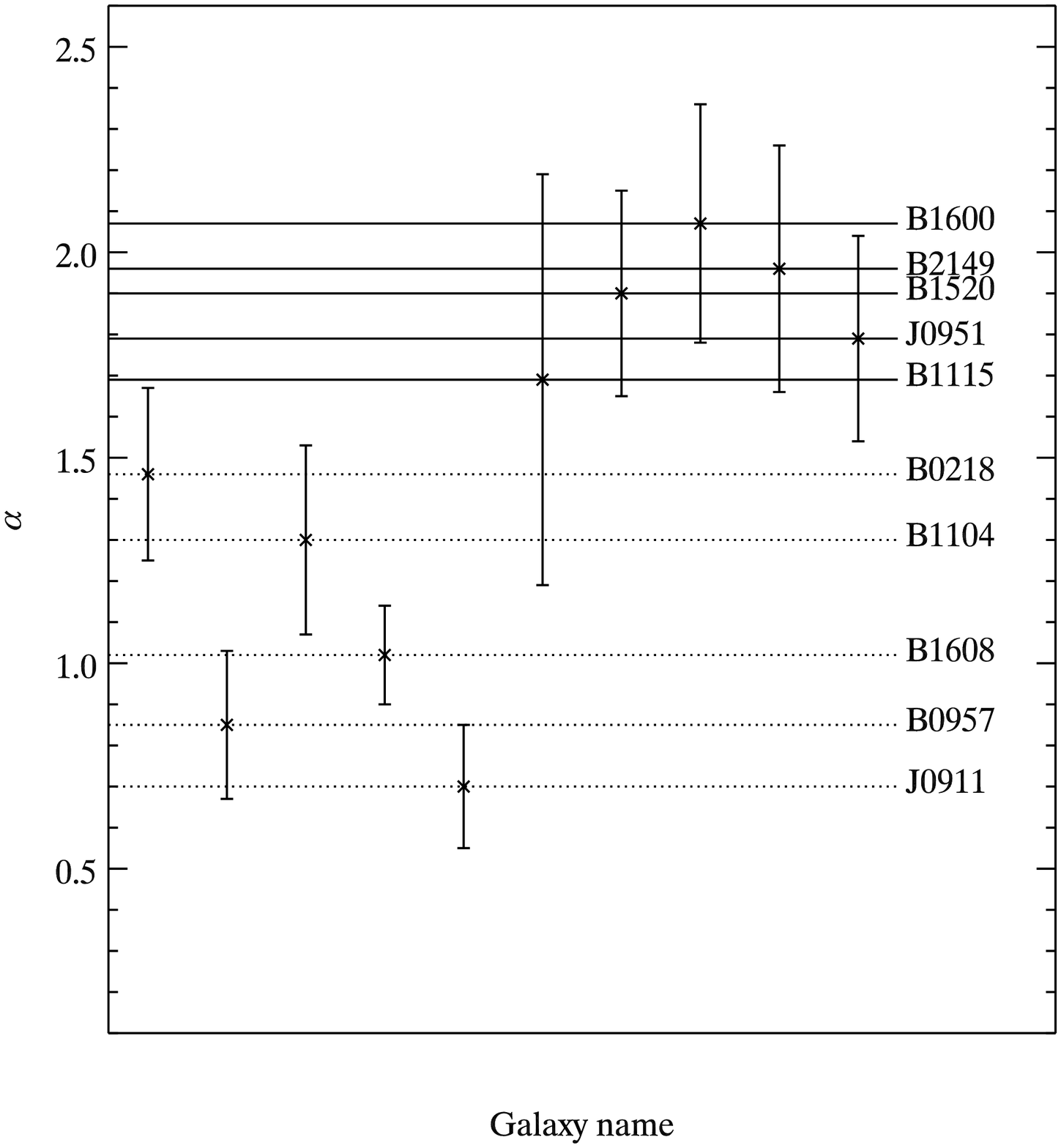}
\caption{The mean steepness, $\alpha$ (where $\rho\propto r^{-\alpha}$),
for the ten lensing galaxies. Formal errors are marked.}
\label{fig:galenviron}
\end{center}
\end{figure}

The five systems close to $r^{-2}$ are all relatively isolated
galaxies (though nearby galaxies contributing some external shear are
often present).  From estimates of the stellar mass through population
synthesis models \citep{2005ApJ...623L...5F}, all five have both stars
and dark matter contributing to lensing.  In other words, these are
all galaxies where we would expect an isothermal mass distribution.

The five shallower lenses are different.  In B0957+561 and J0911+055
the main lensing galaxy is in a cluster, while B0218+357 and B1104+180
have little starlight, suggesting a dark-matter dominated lens; in all
of these cases we would expect a shallower profile than $r^{-2}$.  Finally, in
B1608+656 the lens consists of two interacting galaxies; here the
density peak of the host galaxy will be averaged with the lower
density material from the infalling system, hence here again we would
expect a shallower profile than $r^{-2}$.

It is worth adding an extra comment for B0218+357, which appears distinct from the other lenses. Our five steep lenses have the outer-most image at a projected radius of $<10$\,kpc. By contrast, all of the shallow lenses have the outer-most image at $>10$\,kpc, with the exception of B0218+357. B0218+357 has its outermost image at a projected radius of little over $\sim 2$\,kpc, while its profile is of intermediate steepness, with $\alpha \sim 1.5$. It may be that these peculiarities are the result of a poor determination of the optical centre for this lens, due its small size on the sky \citep{2003MNRAS.338..599B}. Such errors are not currently taken into account in our models. We hope to investigate this lens in more detail in future work. 

We conclude that while the inferred steepness has a large uncertainty,
the identification of nearly $r^{-2}$ versus shallower profiles is a
strong result.

\section{Comparison with previous work}\label{sec:pastwork}

Some recent work has concluded that if measured time delays are to be
consistent with $H_0^{-1}\simeq 14\Gyr$ ($H_0\simeq 70\legacy$) then
the galaxy mass profiles involved must be significantly steeper than
$r^{-2}$.  In particular, \cite{2004mmu..symp..117K} find that
isothermal lenses give $H_0^{-1}=20\pm1\Gyr$ ($H_0=48\pm3\legacy$)
from the measured time delays.  \cite{2006A&A...460..647D} report that
$H_0=71\legacy$ is consistent with an $r^{-2}$ profile within
$3\sigma$. However, when they separate the lens sample into doubles
and quads, the $3\sigma$ uncertainties do not overlap, so it is not
clear how to interpret their result.  \cite{2006ApJ...649..599K}
attempt to break the steepness degeneracy using a single velocity
dispersion for each galaxy.  They conclude that all of their sample
(time delays are not available for those lenses) is consistent within
observational errors with isothermal profiles. However, when the same
method is applied to the time-delay lens B1115+080
\citep{2002MNRAS.337L...6T} a profile significantly steeper than
isothermal is inferred.  See \citet{2007astro.ph..2741D} for a recent
discussion of the difficulties and proposed solutions.

\begin{figure}[t]
\begin{center}
\includegraphics[width=0.49\textwidth]{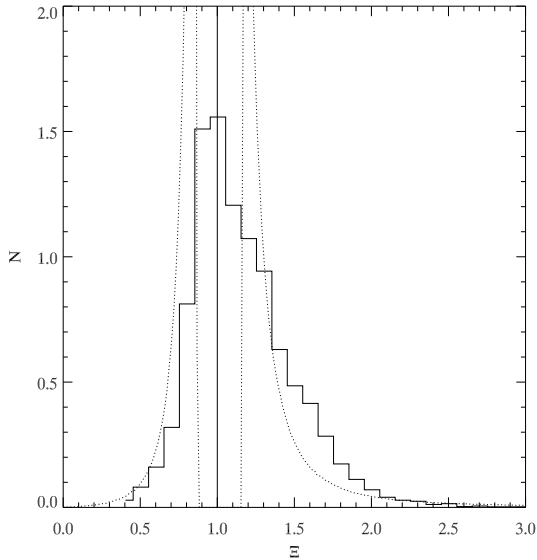}
\caption{Distribution of reduced time delays, $\Xi$ [see equation
  \ref{Xi}], for the simulated galaxy lens (solid histogram). To
  produce this plot we traced rays from a uniform distribution of
  sources in the source-plane and calculated the distribution of
  arrival times for their images in the image plane. Over-plotted are
  the similar results for an isothermal lens (solid vertical line) and
  a first-order perturbed isothermal lens (Equation \ref{eqn:pertans},
  with $\epsilon= 0.15\theta_\mathrm{E}$; dotted line). The latter two
  are calculated analytically in section \ref{sec:perturb}.  Since an
  isothermal lens has $\Xi=1$, this plot measures the deviation of
  time delays from isothermal.}
\label{fig:taudist}
\end{center}
\end{figure}

Somewhat in contrast is the work of \cite{2005ApJ...623L...5F}, who
compared stellar-mass profiles derived from population synthesis models with
total-mass profiles reconstructed using {\em PixeLens\/} for 18
early-type lensing galaxies, of which six had time-delay measurements
(all also used in the present work).  Although that paper does not
discuss the steepness of the profiles as such, it finds no inconsistencies
when the measured time-delays were imposed as constraints along with
$H_0^{-1}=14\Gyr$.  Plausible dark-matter fractions are inferred for
all the lenses, and the tilt of the fundamental plane is reproduced.

The above studies all use different (though sometimes overlapping)
samples of lenses.  This is not a problem for pixelated models, since
the method explicitly allows for heterogeneous samples.  However, in
studies that assume a simple form for lensing galaxies, sample
selection becomes an issue and may be partly responsible for
differences between results for $H_0$ and density profiles presented
in the literature so far.  However, as we shall now demonstrate, the
details of the mass model have a far more important effect.

Figure \ref{fig:taudist} shows the distribution of reduced time delays $\Xi$ [see Equation~\ref{Xi}] for the simulated lens with a uniform distribution of sources (solid histogram).  Although the simulated lens is well-fit by an isothermal density distribution (Figure \ref{fig:nbodyprofile}),
its distribution of reduced time delays is not.

\section{Time delays in a perturbed isothermal lens}\label{sec:perturb}

The unexpectedly broad histogram in Figure~\ref{fig:taudist} is
readily understood from a perturbative calculation, as we now show.

Consider an isothermal lens with Einstein radius $\theta_\mathrm{E}$,
with a source at $(p,0)$.  As is well known, the arrival time at sky
position $(x,y)$ is
\begin{equation}
\tau(x,y) = {\textstyle\frac12} (x-p)^2 + {\textstyle\frac12} y^2 
          - \theta_\mathrm{E} \sqrt{x^2+y^2} .
\label{eqn:arrivaliso}
\end{equation}
Equating the gradient of $\tau$ to zero gives the image positions
(provided $|p|<\theta_\mathrm{E}$)
\begin{equation}
x_{+} = p + \theta_\mathrm{E} , \quad x_{-} = p - \theta_\mathrm{E} ,
\quad y_\pm = 0.
\end{equation}
Here $(x_+,0)$ is a minimum, $(x_-,0)$ is a saddle point, and there is
no third image because the potential is singular.  For $p=0$ the two
images merge into an Einstein ring.

Plugging the image positions back into Equation~(\ref{eqn:arrivaliso}),
we obtain the time delay
\begin{equation}
\tau_{-} - \tau_{+} = 2p \, \theta_\mathrm{E} .
\end{equation}
Note that $\tau_+ < \tau_-$.  It is easily verified that the reduced
time delay (\ref{Xi}) is unity.

Now imagine that we add a perturbing potential $\delta\Phi$ to
Equation~(\ref{eqn:arrivaliso}).  As is usual in perturbation theory,
we Taylor expand about the unperturbed solution.
\begin{equation}
\delta \Phi_{\pm} = \epsilon_{\pm} x + \epsilon'_{\pm} y
+ O(\epsilon^2)
\end{equation}
where $\epsilon_{\pm}$ and $\epsilon'_{\pm}$ are Taylor-expansion
coefficients around the two images $(x_\pm,0)$.  Equating the gradient
of the perturbed arrival time $\tau-\delta\Phi$ to zero, we have to
first order:
\begin{equation}
x_{\pm} = p \pm \theta_\mathrm{E} + \epsilon_{\pm} , \quad
y_\pm = \epsilon'_\pm (1\pm \theta_\mathrm{E}/p) .
\label{eqn:pertimage}
\end{equation}
We now write $\epsilon$ for the mean of $\epsilon_\pm$ and neglect
their difference.  The numerator in the reduced time delay (\ref{Xi})
remains $2p\,\theta_\mathrm{E}$ to leading order, whereas the
denominator is now $2(p+\epsilon)\,\theta_\mathrm{E}$.  Hence we have
\begin{equation}
\Xi = \left(1+\frac{\epsilon}{p}\right)^{-1} .
\label{eqn:pertans}
\end{equation}
Now comes the key point. Since $p$ can be small, $\epsilon/p$ need not
be small.  Hence {\em small perturbations of an isothermal lens can
produce zeroth-order changes in the reduced time delay}.

In Figure~\ref{fig:taudist} we have over-plotted the distribution of
$\Xi$ for $\epsilon=0.15\,\theta_\mathrm{E}$ (dotted line).  For this computation, we
had $\sqrt p$ uniformly distributed, corresponding to a uniform
distribution of sources in a disk.  There is a gap in the perturbed
distribution around $\Xi=1$, which arises because
$|p|<\theta_\mathrm{E}$.  Otherwise, however, our simple first-order
perturbative calculation qualitatively reproduces the distribution for
the simulated galaxy, including the asymmetry.

The sensitivity of the time delays to
perturbations of the lens has been noted in the literature
before. \cite{2000ApJ...544...98W} studied perturbations of the
isothermal lens, and found that a shear perturbation could produce a
large change in the time delays. Similarly, \cite{2006astro.ph..9694O}
considered more general perturbations to the isothermal sphere and
found that perturbations to the isothermal potential can give large
changes in the time delays -- indeed our result is implicit in their Figure 1. However, these works did not stress the importance of comparing perturbations to the time delays against perturbations to the image
positions.  Only \cite{1986ApJ...310..568B} appear to have commented directly 
on this: {\em ``It has been our experience that fairly small changes
in the lensing potential may introduce much larger difference in the
delays than in the image properties.''}

\section{Conclusions}\label{sec:conclusions}

Our main result on density slopes (shown in Figures~\ref{fig:obs1} and
\ref{fig:obs2}, and summarized in Figure \ref{fig:galenviron}) is that
five of the galaxies studied are close to $\rho\propto r^{-2}$,
whereas the other five are shallower and clustered around $\rho \propto r^{-1}$.  The former are all relatively isolated galaxies where stars and dark matter probably both
contributing significantly to lensing.  The shallow systems are
different: B0957+561 and J0911+055 lie in clusters; in B1104--181 the lens images are far out in the halo suggesting that this lens is probably dark-matter dominated; B0218+357 has little starlight, suggesting also a dark-matter dominated lens, though the centroid is difficult to determine for this system; and finally B1608+656 is a pair of interacting
galaxies.  Thus the latter five are just those systems which we might
have guessed would be shallower.  Our technique is tested on a mock
survey of five lenses generated from a galaxy-formation simulation,
from which $H_0$ and the mass profile are both recovered within the
claimed uncertainties (Figures~\ref{fig:nbodyhubble} and
\ref{fig:nbodyprofile}).

A significant finding is that, when calculating time delays:
\begin{equation}
\hbox{a nearly-isothermal {\it mass distribution}} \not\simeq
\hbox{a perfect isothermal lens}.
\label{notimplies}
\end{equation}
We demonstrate this through a simple perturbation calculation showing
how a small deviation from an isothermal potential can be amplified in
the time delays.  We also verify this effect in a simulated galaxy,
where time delays from 0.5 to 2 (and, in rare cases, an arbitrarily large number) times the value expected for a
isothermal lens can occur (Fig.~\ref{fig:taudist}), depending on the
source position.

Although fleetingly anticipated by \cite{1986ApJ...310..568B} and
suggested by some more recent modeling work
\citep{1999AJ....118...14B,2003ApJ...582....2Z,2006ApJ...653..936S},
the assertion (\ref{notimplies}) is contrary to the majority of lens
models so far. Such models assume that lensing time delays are
completely determined by the image positions and radial power-law
index.  Once we drop that invalid approximation, the apparent
contradictions between time delays and mass profiles much discussed in
several recent papers (see Section~\ref{sec:pastwork}) are resolved.

\section{Acknowledgements}
We thank Jonathan Coles for providing the latest version of {\it PixeLens.}

\appendix

\section{Lensing theory: scales and degeneracies}

This paper makes use of a number of rather specialized results from
lensing theory.  These are well-known known to experts, but since the
original derivations are spread over diverse papers and a long time
span, we review them here.

\subsection{The arrival-time surface}

We start with the expression for the change in travel time for a
virtual photon originally from a source at sky position $\bbeta$ but
deflected at the lens so that the observer sees it coming from
sky-position $\btheta$:
\begin{equation}
c\,t(\btheta;\bbeta) = (1+\zl) \frac{\Dl\Ds}{2\Dls} (\btheta-\bbeta)^2
                  - (1+\zl) \frac{4G}{c^2}
\int \Sigma(\thp) \ln |\btheta-\thp| \, d^2\thp .
\label{physarriv}
\end{equation}
This is just Equation~(2.6) from \cite{1986ApJ...310..568B} with all
the dimensions put back in; as in that paper, $\Sigma$ is the
sky-projected density, $\zl$ is the lens redshift, $\Dl$ is the
angular-diameter distance from observer to lens, and so on.  The first
term on the right is the geometrical path difference between a
deflected and undeflected photon trajectory, and the last term is the
general-relativistic delay. The above equation assumes that the lens
is a)~infinitesimally thin (compared to $\Dl$ and $\Dls$), and b)~lies
in a representative patch of the Universe
\citep{1986ApJ...310..568B}. Both of these are conservative
assumptions.

The light-travel time (\ref{physarriv}) can be expressed in a useful
dimensionless form by some substitutions. First, we have the scaled
time variable
\begin{equation}
\tau = \left[(1+\zl) \frac{\Dl\Ds}{\Dls}\right]^{-1} c\,t .
\label{time-scale}
\end{equation}
Each of $\Dl,\Ds,\Dls$ is $(c/H_0)$ times a redshift- and
cosmology-dependent factor of order unity.  In other words
\begin{equation}
\tau \sim H_0^{-1} t
\end{equation}
or the time delay in units of the Hubble time.  Next, we introduce the
dimensionless distance factor
\begin{equation}
\fls = \frac{\Dls}{\Ds} .
\label{zs-scale}
\end{equation}
Then we consider
\begin{equation}
\kappa = \frac{4\pi G}{c^2} \Dl \Sigma
\label{mass-scale}
\end{equation}
which can be interpreted as $\Sigma$ in units of the critical density
for sources at infinity.  (In many papers, the $\fls$ factor is
absorbed inside the definition of $\kappa$; that refers $\kappa$ to a
particular $\zs$.)  Finally, we write an operator that solves
Poisson's equation in two dimensions
\begin{equation}
\nabla^{-2} f \equiv \frac1{2\pi}
            \int f(\thp) \ln |\btheta-\thp| \, d^2\thp .
\end{equation}
With all these substitutions we obtain
\begin{equation}
\tau(\btheta;\bbeta) = {\textstyle\frac12} (\btheta-\bbeta)^2
                - 2 \fls \nabla^{-2} \kappa(\btheta) .
\label{arriv}
\end{equation}
For given $\bbeta$, the abstract surface $\tau(\btheta)$ is called the
arrival-time surface.  It is abstract because real photons do not arrive
from all $\btheta$ --- they arrive only from $\btheta$ that make
$\tau(\btheta)$ extremal (Fermat's principle).  In other words,
images appear where the arrival-time surface has a maximum, minimum,
or saddle-point.

\subsection{Dimensionless time delays}

Equation (\ref{arriv}) is completely dimensionless if we measure
angles in radians, which emphasizes that a lens model
$\kappa(\btheta)$ on its own is dimensionless.  Scales in lensing
enter through equations (\ref{time-scale}), (\ref{zs-scale}) and
(\ref{mass-scale}), which are all model-independent.  \citep[See][for
an interesting discussion of this point.]{1990CurrSci.50.1044}

The last statement may seem paradoxical.  If lens models are
dimensionless, how can a lens model help infer $H_0$?  The resolution
is that lens models relate two time scales ---$H_0^{-1}$ and
observable time delays--- even though they have no time scales of
their own.  Because of this, we can speak of a dimensionless time
delay even for an observed lens with a measured time delay, and this
is very useful for comparing a heterogeneous set of lenses.  Two ways
of turning a measured delay $\Delta t$ into a dimensionless number are
suggested in \cite{2004A&A...414..425S}.

One possibility is to compute
\begin{equation}
\varphi = \frac{\Delta t}{{\textstyle\frac1{16}}(\theta_1+\theta_2)^2D}
\label{varphi}
\end{equation}
where where $\theta_1,\theta_2$ are the lens-centric distances (in
radians) of the first and last images to arrive, $\Delta t$ is the
observed time delay between them, and $D$ is the factor
$(1+\zl)(\Dl\Ds/\Dls)$ from equation (\ref{time-scale}) --- which
recall is $\propto H_0^{-1}$.  The factor
$\frac1{16}(\theta_1+\theta_2)$, with angles in radians, is
approximately (exactly for an isothermal lens) the fraction of the sky
covered by the Einstein ring.  It turns out that $\varphi$ lies in the
range $\simeq [0,2]$ for doubles and $\simeq [2,8]$ for quads.  This
property was exploited by \cite{2006ApJ...650L..17S} to compare
observed and $N$-body time delays.

Another possibility is the reduced time delay
\begin{equation}
\Xi = \frac{\Delta t}{{\textstyle\frac12}(\theta_1^2-\theta_2^2)D}
\label{Xi}
\end{equation}
which has a wider range than $\varphi$ but has the attractive property
that for an isothermal lens $\Xi = 1$.  \cite{2006astro.ph..9694O}
uses $\Xi$ as a basis to compare the distributions of observed and
model time delays; we also use this quantity in section \ref{sec:perturb}.

\subsection{Steepness and other degeneracies}

Let us rewrite the arrival time slightly as
\begin{equation}
\tau = 2\,\nabla^{-2} (1 - \fls\kappa) - \btheta\cdot\bbeta
\label{arriv2}
\end{equation}
where we have used the fact that in two dimensions
$\nabla^2\theta^2=4$, and also discarded a $\bbeta^2$ term since it
has no optical effect.  Now consider the rescaling
\begin{equation}
\tau' = \lambda \tau , \quad
(1-\fls\kappa') = \lambda (1-\fls\kappa), \quad
\bbeta' = \lambda \bbeta .
\label{rescale}
\end{equation}
applied to Equation (\ref{arriv2}), where $\lambda$ is constant in the
region of images.  There is no effect on the image positions or
relative magnifications --- basically we are redefining the contour
spacing in a contour map of the arrival-time surface (such as
Blandford \& Narayan's Figure~2) without altering the figure.
However, time delays change because $\tau$ is rescaled, while the
total magnification changes because $\bbeta$ is rescaled.  Meanwhile,
rescaling $(1-\fls\kappa)$ means that the mass profile gets steeper
(if $\lambda>1$) or shallower (if $\lambda<1$).

Equation (\ref{rescale}) is the well-known steepness
degeneracy.\footnote{It is often called the mass-sheet degeneracy,
because one of its limits is a mass sheet $\fls\kappa'=1$.
Unfortunately the term ``mass-sheet'' easily leads to a misconception
that the degeneracy consists of {\em adding\/} a mass sheet.  Hence
the name ``steepness degeneracy'' is preferable.} The simple
geometric derivation given here is from \cite{2000AJ....120.1654S},
but it was first derived by a different method by
\cite{1985ApJ...289L...1F}.

The steepness degeneracy is broken if there is a range of $\zs$
\citep{1998AJ....116.1541A,2006ApJ...652L...5S}.  That is because a
range of $\zs$ gives a range of $\fls$, and Equation (\ref{rescale})
can only be applied if $\fls$ is constant.  Another way to break the
degeneracy is to have number counts of weakly lensed objects
\citep{2002A&A...386...12D} because then the total magnification is
constrained and $\bbeta$ cannot be freely rescaled.  However for
galaxy lenses, neither of these routes is in practice available.
Multiple sources at the same $\zs$ do not help here, since the
rescaling (\ref{rescale}) can be applied to multiple sources, so even
an Einstein ring does not break the degeneracy
\citep{2001AJ....122..585S}.

That the steepness degeneracy is a serious problem for galaxy lenses
is now widely appreciated, and researchers agree that steeper mass
profiles lead to larger values of $(H_0\,\Delta t)$.  But as our
Figure~\ref{fig:taudist} shows, significant degeneracies can persist
even at fixed slope.  These degeneracies appear to consist of
rescalings of the type (\ref{rescale}) but with varying $\lambda$.
The details remain to be investigated.

\bibliographystyle{apj}
\bibliography{/Users/justinread/Documents/LaTeX/BibTeX/refs}

\end{document}